\documentstyle[aps,prl,epsf,multicol]{revtex}

\newcommand{\be}{\begin{equation}}
\newcommand{\ee}{\end{equation}}
\newcommand{\ba}{\begin{eqnarray}}
\newcommand{\ea}{\end{eqnarray}}

\def\opone{\leavevmode\hbox{\small1\kern-3.8pt\normalsize1}}
\def\l{\label}
\def\c{\cite}
\def\r{\ref}
\def\la{\langle}
\def\ra{\rangle}
\def\e{{\rm e}}
\def\i{{\rm i}}
\def\f{\frac}

\font\eightln=line10 at8pt
\catcode`\@=11
\def\doublearrow{\@ifnextchar [{\@doublearrow }{\@doublearrow[0]}}
\def\@doublearrow[#1]{\mathrel{\,\lower0.15ex
  \hbox{\let\@linefnt\eightln\unitlength0.6ex\begin{picture}(4,3)
  \ifcase#1\put(4,0.8){\vector(-1,0){4}}\put(0,2){\vector(1,0){4}}
  \or      \put(0,0.8){\vector(1,0){4}}\put(0,2){\vector(1,0){4}}
  \or      \put(4,0.4){\vector(-2,1){4}}\put(0,0.4){\vector(2,1){4}} 
  \fi
  \end{picture}}\,}}
\catcode`\@=12

\font\eightln=line10 at8pt
\catcode`\@=11
\def\triplearrow{\@ifnextchar [{\@triplearrow }{\@triplearrow[0]}}
\def\@triplearrow[#1]{\mathrel{\,\lower0.15ex
  \hbox{\let\@linefnt\eightln\unitlength0.6ex\begin{picture}(4,3)
  \ifcase#1\put(0,0.3){\vector(1,0){4}}\put(0,1.5){\vector(1,0){4}}
           \put(0,2.7){\vector(1,0){4}}
    \or\put(4,0.3){\vector(-1,0){4}}\put(4,1.5){\vector(-1,0){4}}
           \put(0,2.7){\vector(1,0){4}}
    \or\put(4,2){\vector(-2,-1){4}}\put(4,3){\vector(-2,-1){4}}
           \put(0,3){\vector(4,-3){4}}
    \or\put(4,0){\vector(-1,0){4}}\put(4,3){\vector(-2,-1){4}}
           \put(0,3){\vector(2,-1){4}}
    \or\put(0,0){\vector(2,1){4}}\put(4,3){\vector(-2,-1){4}}\put(4,0){\vector(-4,3){4}}
  \fi
  \end{picture}}\,}}
\catcode`\@=12

\font\eightln=line10 at8pt
\catcode`\@=11
\def\tripline{\@ifnextchar [{\@tripline }{\@tripline[0]}}
\def\@tripline[#1]{\mathrel{\,\lower0.15ex
  \hbox{\let\@linefnt\eightln\unitlength0.6ex\begin{picture}(4,3)
  \ifcase#1\put(0,0.3){\line(1,0){4}}\put(0,1.5){\line(1,0){4}}
           \put(0,2.7){\line(1,0){4}}
  \or\put(0,0){\line(2,1){4}}\put(0,1){\line(2,1){4}}
           \put(0,3){\line(4,-3){4}}
  \or\put(0,3){\line(2,-1){4}}\put(0,2){\line(2,-1){4}}
           \put(0,0){\line(4,3){4}}
  \or\put(0,0){\line(4,3){4}}\put(0,1.5){\line(1,0){4.5}}
           \put(0,3){\line(4,-3){4}}
  \or\put(0,0.3){\line(1,0){4}}\put(0,1){\line(2,1){4}}
           \put(0,3){\line(2,-1){4}}
  \or\put(0,0){\line(2,1){4}}\put(0,2){\line(2,-1){4}}
           \put(0,2.7){\line(1,0){4}}
  \fi 
 \end{picture}}\,}}
\catcode`\@=12

\begin{document}

\title{Universal spectral form factor for chaotic dynamics}

\author{Stefan Heusler$^1$, Sebastian M\"uller$^1$, Petr Braun$^{1,2}$, and Fritz Haake$^1$}

\address{$^1$Fachbereich Physik, Universit\"at Duisburg-Essen,
45117 Essen, Germany\\
$^2$Institute of Physics, Saint-Petersburg University, Saint-Petersburg 198504, Russia}

\maketitle

\begin{abstract}

We consider the semiclassical limit of the spectral form factor $K(\tau)$ of
fully chaotic dynamics. Starting from the
Gutzwiller type double sum over classical  periodic orbits we set out to
recover the universal behavior predicted by random-matrix theory, both for
dynamics with and without time reversal invariance. For times smaller than half
the Heisenberg time $T_H\propto \hbar^{-f+1}$, we extend the previously known
$\tau$-expansion to include the cubic term. Beyond confirming random-matrix
behavior of individual spectra, the virtue of that extension is that the ``diagrammatic rules''
come in sight which determine the families of orbit pairs responsible for
all orders of the $\tau$-expansion.

\end{abstract}
\begin{multicols}{2}

{\it Introduction:} One of the fascinating quantum signatures of chaos is universal
behavior of the correlation functions of the spectral density of energy levels, for
general hyperbolic dynamics \c{BGS}.
Three universality classes were suggested by Dyson and Wigner; one, called
``unitary'', has no time reversal symmetry, while the other two do have Hamiltonians
$H$ commuting with an anti-unitary time reversal operator ${\cal T}$; if ${\cal T}^2=1$ 
one speaks of the ``orthogonal''
class while the ``symplectic'' case has ${\cal T}^2=-1$. The Fourier transform of the 
two-point correlator of the level density, called spectral
form factor, is predicted by random-matrix theory (RMT) \c{Bibel} as
\be
\l{1}
K_{\rm uni}(\tau)=\tau\;,\quad K_{\rm orth}(\tau)=2\tau-\tau\ln(1+2\tau)\,,
\ee
in the unitary and the orthogonal case; here $\tau$ is a time
made dimensionless by referral to the Heisenberg time $T_H\propto\hbar^{-f+1}$, with $f$ the 
number of freedoms; the results (\r{1}) hold in the (semiclassical) limit of large dimension of the
matrix representation of $H$ and for times up to the Heisenberg time,
$0\leq\tau\leq 1$. The orthogonal form factor allows for the Taylor expansion
$K_{\rm orth}(\tau)=2\tau-2\tau^2+2\tau^3+\dots$, for $0\leq\tau<\f{1}{2}$.

Understanding the observed fidelity of individual dynamics to RMT has been an elusive goal, in spite 
of considerable efforts based on parametric level dynamics, semiclassical periodic-orbit theory,
and the so-called non-linear sigma model \c{Bibel}. We shall here take a non-trivial 
step towards that goal, on semiclassical ground.

Periodic-orbit theory \`a la Gutzwiller \c{Gutzi,Bibel} gives the form factor of an
individual spectrum as the double sum
\be
\l{2}
K_{\rm po}(\tau)=\left\la\sum_{\gamma,\gamma'}A_\gamma A_{\gamma'}^*
\e^{\i(S_\gamma-S_{\gamma'})/\hbar}\delta\big(\tau-\textstyle{\f{T_\gamma+
T_{\gamma'}}{2T_H}}
\big)\right\ra
\ee
where $S_\gamma,T_\gamma$, and $A_\gamma$ are the classical action (including the
Maslov phase), period, and stability amplitude of the $\gamma$th orbit; the angular brackets 
demand averages over (i) the center energy $(E+E')/2$ in the product $\rho(E)\rho(E')$ of two level
densities before doing the Fourier transform w.r.t. the energy difference
$E-E'$ and (ii) over a time interval small compared to the Heisenberg time.
Most orbit pairs $\gamma,\gamma'$ interfere destructively in the double sum (\r{2}).
Finite contributions arise only from families of pairs wherein the
action difference $S_\gamma-S_{\gamma'}$ can be continuously steered through the
quantum scale $\hbar$ toward zero, by varying  parameters defining the family (for early premonitions 
see \c{Smiley}). Periods and stability amplitudes do not differ noticeably  within such an orbit pair.

Berry's ``diagonal approximation'' \c{Berry} includes the trivial pairs
$\{\gamma,\gamma\}$ and, given $\cal T$ invariance,
$\{\gamma,\bar{\gamma}\}$ where the overbar indicates time reversal; it yields
$K^{(1)}=\beta\sum_\gamma|A_\gamma|^2\delta(\tau-\f{T_\gamma}{T_H})=
\beta\tau$ where $\beta=1$ without and $\beta=2$ with $\cal T$ invariance, due to the doubling of 
contributing pairs in the latter case. The 
sum $\sum_\gamma|A_\gamma|^2\delta(\tau-\f{T_\gamma}{T_H})=
\tau$, known as the sum rule of Hannay and Ozorio de Almeida (HOdA)\c{HOdA}, reflects
ergodicity for long periodic orbits. In view of
(\r{1}) the diagonal approximation gives $K_{\rm uni}$ in full, and the first
term of the $\tau$-expansion of $K_{\rm orth}$.

Sieber and Richter \c{Sieber} recently found a family of orbit pairs which for the
$\cal T$ invariant Hadamard-Gutzwiller model yields the quadratic term of the $\tau$-expansion, 
$K^{(2)}_{\rm orth}=-2\tau^2$, as in (\r{1}). Each Sieber-Richter (SR) pair has a close self-encounter 
which in configuration space
looks like a small-angle crossing for one orbit and like a narrowly avoided crossing
for the partner orbit. Generalizations to arbitrary hyperbolic systems with two freedoms were given 
in \c{Mueller,Spehner,Regensburg} and for more freedoms in \c{RegEss}. 

Before identifying the new families of orbit pairs giving 
$K^{(3)}_{\rm orth}=2\tau^3,\, K^{(3)}_{\rm uni}=0$ we must briefly review SR. 

Each self-encounter involves two orbit stretches which are nearly
mutually time reversed; it may be depicted as
$\doublearrow$ (or \,$\doublearrow[2]$),
arrows indicating sense of traversal. On either side of the ``encounter
graph'' $\doublearrow$, each of the two orbits has a
long loop  attached. Assuming symbolic dynamics available (to uniquely define periodic orbits and 
even, approximately, short orbit stretches by symbol sequences; our results are valid more
generally) we could write $E$ and $\bar{E}$ for the
two (nearly) mutually time reversed orbit stretches in the encounter region, $R,L$ for the two
long loops, and $\bar{L}$ for the time reversed of $L$; we may thus write $ER\bar{E}L$ for an orbit and
$ER\bar{E}\bar{L},\,LE\bar{R}\bar{E}$ for its SR partners\c{BHMH,Mueller}. Note that the orbits in a SR pair 
traverse one loop in the same sense while the senses of traversal are opposite for the other loop; 
here, $\cal T$ invariance is seen as required for SR pairs to exist. 

Following \c{BHH,Spehner,Regensburg} we
parametrize an encounter with the help of a surface of  section $\cal P$ transverse to an orbit,
say $\gamma=ER\bar{E}L$, somewhere within the encounter; $\cal P$ is two dimensional for $f=2$, the case we limit ourselves to. The stretch $E$ 
pierces through $\cal P$ in a point $x_a$ which can be made the origin of a
coordinate system spanned by  tangent vectors $\hat{e}_s$ and $\hat{e}_u$ to the
stable and unstable manifolds of $\gamma$ through $x_a$ in $\cal P$. A second piercing is associated 
with $\bar{E}$ and thus opposite in sense; it happens after the traversal 
of the right loop at a point $x_b$. We define a close-encounter region by requiring the unstable and stable components of the difference ${\cal T}x_b-x_a$ to respect a classically small bound $c$ independent of $\hbar$, 
\be
\l{3}
{\cal T}x_b-x_a=u \hat{e}_u+s \hat{e}_s\,, \quad |u|\leq c\,,\quad |s|\leq c\,.
\ee
Moving $\cal P$
we leave the encounter after a time $t_u$ given (asymptotically) by $|u|\e^{\lambda t_u}=c$. 
Conversely, we find the start of the encounter going backwards in time by $t_s$ with  
$|s|\e^{\lambda t_s}=c$; here $\lambda$ is the Lyapunov exponent of the system. The duration of 
the encounter thus is
\be
\l{4}
t_{\rm enc}=t_s+t_u=\f{1}{\lambda}\ln\f{c^2}{|us|}\,.
\ee
Roughly, linearization
of the dynamics about any point within the encounter breaks down at either end.

As was shown in \c{BHH,Mueller,Spehner,Regensburg} the partner orbit $ER\bar{E}\bar{L}$ pierces through
$\cal P$ first in $x_a^p= x_a+u\hat{e}_u$ and then in $x^p_b$ with
${\cal T}x_b^p= x_a+s\hat{e}_s$. Moreover, the orbits in a SR pair differ in action by the area of the parallelogram spanned by the four points
$x_a,x_a^p,{\cal T}x_b,{\cal T}x_b^p$ \c{Spehner,Regensburg},
\be
\l{5}
\Delta S=us\,;
\ee
that product is canonically invariant and thus independent of the precise location of the surface 
$\cal P$ \big(The vectors $\hat{e}_s, \hat{e}_u$ are pairwise normalized as 
$\hat{e}_s\wedge \hat{e}_u=1$\big). Inasmuch as weighty 
contributions to the form factor must have $\Delta S={\cal O}(\hbar)$, we can conclude that the 
duration of relevant encounters has the order of the
Ehrenfest time $T_{E}=\f{1}{\lambda}\ln\f{c^2}{\hbar}$, much
smaller than the period $T={\cal O}(T_H)$.

To evaluate $K^{(2)}(\tau)$ 
we need the cumulative duration $P(u,s|T)duds$
of all orbit stretches within self-encounters of a long orbit $\gamma$ of period $T={\cal O}(T_H)$, with unstable and 
stable components of ${\cal T}x_b-x_a$  in  $[u,u+du]$ and $[s,s+ds]$.
Ergodicity yields (Refs. \c{Mueller,Spehner,Regensburg} use different conventions)
\be
\l{6}
P(u,s|T)duds=T(T-2t_{\rm enc})\Omega^{-1}duds
\ee
with $\Omega$ the volume of the energy shell. This results from integrating the ergodic return 
probability density $\Omega^{-1}$ over the two times of piercing of the orbit through a section $\cal P$. 
The factor $T$ indicates that one piercing, say the one at $x_a$, may occur at any time in the interval 
$[0,T]$. The time of the subsequent piercing at $x_b$ can then lie only in an interval
of length $T-2t_{\rm enc}$, hence the second factor in (\r{6}); this is because both traversals of the 
encounter region have length $t_{\rm enc}$ and may not overlap. (Overlapping stretches $E,\bar{E}$ are 
either impossible, as in the Hadamard-Gutzwiller model \c{BHMH}, or indicate an orbit with a 
self-retracing loop identical with its SR partner \c{Mueller}). Note that in the density $P(u,s|T)$ each 
encounter is weighted with the duration ${\cal N}t_{\rm enc}$ ;  the combinatorial factor ${\cal N}=2$ arises since in $\gamma= ER\bar{E}L$ the two stretches $E,\bar{E}$ are equivalent; we must therefore employ $\f{P(u,s|T)}{{\cal N}t_{\rm enc}}$, to count each encounter only once \c{foot}. 

The contribution to the form factor reads $K^{(2)}_{\rm orth}=\\
\big(\sum_\gamma|A_\gamma|^2 \delta(\tau-\f{T}{T_H})\big)\int_{-c}^cduds\f{1}{2t_{\rm enc}}
P(u,s|T)2\cos(us/\hbar)$. The sum 
$\big(\sum_\gamma\ldots\big)$ gives the factor $\tau$ through the HOdA sum rule, as for $K^{(1)}$. 
The integral gets no contribution from the leading term $T^2/\Omega$ in $P$,
and this is why the ${\cal O}(T_E/T_H)$ correction in (\r{6}) is important; the integral with
$P\to-2t_{\rm enc}T/\Omega$ becomes independent of the bound $c$ in the
limit $\hbar\to0$ and is easily found (since $t_{\rm enc}$ cancels from the integrand and 
$T_H=\Omega/2\pi\hbar$) as $-2\tau$. The RMT result $K^{(2)}=-2\tau^2$ is thus recovered.

One might wonder why no ``parallel'' encounters with graphs $\doublearrow[1]$
come into play.  The simple reason is that the would-be SR partner of an orbit $EREL$ decomposes into a 
``pseudo-orbit'' (separate periodic orbits $ER$ and $EL$), not admitted to the 
Gutzwiller sum in the first place. However, parallel encounters will be met with below. The 
quantification
(\r{3}) of ``close'' must then be changed to 
$x_b-x_a=u \hat{e}_u+s \hat{e}_s\,,\quad |u|\leq c\,,\quad |s|\leq c$.

The foregoing review of SR has highlighted those twists of the original formulation of \c{Sieber}, in 
particular the appearance of  $\f{P(u,s|T)}{2t_{\rm enc}}$, which make the extension to higher orders 
in $\tau$ a  rather elegant travail, to which we now turn.

{\it Orbits pairs contributing in third order:} We now present five families of orbit pairs relevant for $\tau^3$. They
can be constructed from two SR ``switches'' $\doublearrow$/$\doublearrow[2]$. Three families 
have two separate encounters wherein the orbits differ and four intervening ``loops'' near identical for both orbits; 
two more families arise as one of the loops shrinks, to let the two encounters overlap, see Fig. 1. 

Starting with two independent encounters, we must obviously check three possibilities: ($aa$) both 
``antiparallel'', pictorially
$\;\doublearrow\;\;\doublearrow$\,, ($ap$) antiparallell and  parallel, i.e.
$\;\doublearrow\;\;\doublearrow[1]$\,, and ($pp$) both parallel, i.e.
$\;\doublearrow[1]\;\;\doublearrow[1]$\,. For\\ all of them there are
two distinct ways of filling in intervening lines, with the two encounters either in series ($s$)
or intertwined ($i$). In symbolic notation, those two ways read for
case ($aas$) $E_1A\bar{E_1}BE_2C\bar{E_2}D$ (in series) and for case ($aai$) 
$E_1AE_2B\bar{E_1}C\bar{E_2}D$ (intertwined); here $E_1,\bar{E}_1$  ($E_2,\bar{E}_2$)
represent the two orbit stretches of the
first (second) antiparallel encounter which are nearly mutually time reversed, while the
remaining symbols refer to intervening lines; one immediately checks that only the $aas$
orbit allows for a partner orbit, $E_1\bar{A}\bar{E}_1BE_2\bar{C}\bar{E}_2D$, while the would-be partner 
of the $aai$ orbit turns out a pseudo-orbit decomposing into the two periodic orbits
$E_1AE_2\bar{C}$ and $E_1\bar{B}\bar{E_2}D$. In the same vein we find that for
($ap$) only the $api$ orbit and for ($pp$) only the 
$ppi$ orbit have non-decomposing partners. We have thus identified three families of orbit pairs
with two far apart self-encounters. Clearly, $ppi$ pairs arise 
even without $\cal T$ invariance, while $aas$ and $api$ pairs do require that symmetry.

\begin{figure}
\begin{center}
\vspace{-0.45cm}
\leavevmode
\epsfxsize=0.46
\textwidth
\epsffile{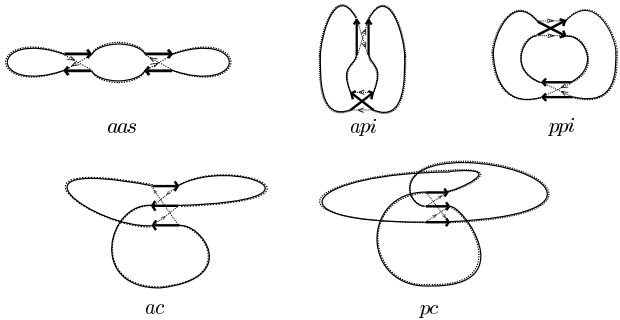}
\end{center}
\caption{The five orbit pairs entering $\tau^3$. Labels describe 
encounters as $a$ntiparallel or $a$ntiunitary-symmetry-required,
$p$arallel, $i$ntertwined, 
$s$erial, and $c$loverleaf.}
\end{figure}

Now on to the families resulting from shrinking away one of the four loops in
the foregoing families. The three remaining loops make for
a cloverleaf ($c$) structure; the encounter region accommodates a triple of oriented short
(${\cal O}(T_{E})$) stretches with encounter graphs
$\;\triplearrow$\,$(pc)$ and $\;\triplearrow[1]$\,(ac); in symbolic notation, they involve $E,E,E$ and
$E,\bar{E},\bar{E}$, respectively. Schematically, the two types of cloverleaf orbits look like the 
thick lines in Fig. 1$ac,pc$. We shall argue that each has a unique partner, shown as dashed 
lines in Fig. 1. The family  $(pc)$ with three parallel 
orbit stretches $\;\triplearrow$\, in the encounter does not require $\cal T$ invariance while the
$\;\triplearrow[1]$\, family $(ac)$ does. 

To check the uniqueness of the partner for $ac$ and $pc$ we start from the respective
thick lines in Fig. 1. Each candidate for partnership must have its three long loops nearly identical 
with those of the starting orbit, save possibly for time reversal; those loops are
differently connected in the encounter region. A sextet of cloverleaf
orbits thus seems to arise, whose encounter graphs are (without arrows, momentarily)\,
$\tripline$\, $\tripline[1]$\, $\tripline[2]$\, $\tripline[3]$\, $\tripline[4]$\, $\tripline[5]$\,.
The last three immediately drop from candidacy since they represent SR partners of $\tripline\,$ 
contributing to $\tau^2$ or entail decomposing pseudo-orbits. Now putting arrows on $\tripline$, say for 
$ac$ as $\triplearrow[1]$ and hooking
on the three loops as in the thick line of Fig. 1$ac$, we find only $\tripline[1]$ to lead to a 
non-decomposing partner, the dashed line in Fig.1$ac$ with the encounter graph $\triplearrow[4] $ 
(plus, of course its time reverse); the uniqueness of the $pc$ partner (up to time reversal, if 
$\cal T$ invariance holds) is shown similarly.    

{\it Duration and action differences:} For the independent-encounter pairs $aas,api,ppi$ the action differences of 
the two encounters sum up to $\Delta S= u_1s_1+u_2s_2$. The triple encounters $ac,pc$ require 
extra thought. 

Beginning with $ac$ we employ
a surface of section ${\cal P}$ through the encounter region and denote the three points of piercing
by $x_{a,b,c}$. As in (\r{3}) we have ${\cal T}x_b-x_a=u_b\hat{e}_u+s_b\hat{e}_s,
{\cal T}x_c-x_a=u_c\hat{e}_u+s_c\hat{e}_s$; the bound $c$ must be respected by all six distances 
$|u_b|,|s_b|,|u_c|,|s_c|,|u_b-u_c|\equiv|u_{bc}|,|s_b-s_c|\equiv s_{bc}$ and the cloverleaf encounter 
lasts for 
\be
\l{7}
t_{\rm enc}^{\rm cl}=\f{1}{\lambda}\ln\f{c^2}{\max\{|u_b|,|u_c|,|u_{bc}|\}
\max\{|s_b|,|s_c|,|s_{bc}|\}}.
\ee
Again, relevant encounters will have durations of the order of the Ehrenfest time 
$t_E=\f{1}{\lambda}\ln\f{c^2}{\hbar}$. 

Now look at an $ac$ pair $\{\gamma,\gamma^p\}$ with encounter graphs $\,\triplearrow[1]$ and 
$\,\triplearrow[4]$\,. To find the action
difference we proceed in two steps. In the first, we employ an auxiliary orbit, the SR partner
$\gamma'$ of $\gamma$ related to the encounter of the ``upper two'' stretches, labelled $a,b$, leaving 
the ``lowest'' stretch $c$ as in $\gamma$. For
$\gamma'$ the encounter region is $\;\triplearrow[3]$\,. According to what we have said above
about the four points of piercing through the surface of section of an SR encounter the piercings of 
$\gamma'$ occur at $x_a'= x_a+u_b\hat{e}_u$ and $x_b'$ with ${\cal T}x_b'=x_a+s_b\hat{e}_s$; 
moreover, the action difference within $\{\gamma,\gamma'\}$ is given by (\r{5}) as 
$S_{\gamma'}-S_{\gamma}=u_bs_b$.
In our second step we arrive at the $ac$ partner $\gamma^p$ of $\gamma$ as the SR partner of $\gamma'$
w.r.t. the stretches $a,c$. Once again invoking what we know about the four piercings in an SR
encounter we have $x_a^p=x_a'+(u_c-u_b)\hat{e}_u$ and ${\cal T}x_c^p=x_a'+s_c\hat{e}_s$, and the
action difference $S_{\gamma^p}-S_{\gamma'}=(u_c-u_b)s_c$. The action difference of the $ac$ pair (valid
also for $pc$) thus reads
\be
\l{8}
\Delta S=S_{\gamma^p}-S_\gamma=u_bs_b+u_cs_c-u_bs_c\,.
\ee
The term $u_bs_c$ represents an ``interaction''. 

{\it $\tau^3$-contributions from two simple encounters:} We first generalize the density (\r{6}) to $P(u_1,s_1,u_2,s_2|T)$, 
(up to the factor $du_1ds_1du_2ds_2$) 
the ``area'' of times $(t_a^1,t_a^2)\in[0,T]^2$ such 
that the points $\{x_a^\mu=x(t_a^\mu),\,\mu=1,2\}$ of piercing are followed by piercings at 
$x_b^\mu=x(t_b^\mu)$ with unstable and stable components of $x_b^\mu-x_a^\mu$ (for parallel encounters) 
and  ${\cal T}x_b^\mu-x_a^\mu$ (for antiparallel encounters) in the intervals 
$[u_\mu,u_\mu+du_\mu],\,[s_\mu,s_\mu+ds_\mu]$. We obtain  
\be
\l{9}
P(u_1,s_1,u_2,s_2|T)=\f{1}{6\Omega^2}T\big(T-2(t_{\rm enc1}+t_{\rm enc2})\big)^3\,,
\ee
by integrating the ergodic probability density $\Omega^{-2}$ for two encounters over the four times 
$\{t_a^\mu,t_b^\mu,\, \mu=1,2\}$, respecting the order of those times dictated by the ``diagrams'' 
$aas,api,ppi$ and the general rule that an orbit must leave one encounter region before reentering or 
entering the next one; the latter rule in fact separates  independent-encounter from  cloverleaf families. 
The restrictions on the times of piercing in question give rise to the small but decisive
corrections $t_{{\rm enc},\mu}$ in (\r{9}), as before in (\r{6}). 

Again in analogy with (\r{6}), the density (\r{9}) overcounts a pair of simple encounters, by a factor 
which we shall argue to be ${\cal N}t_{\rm enc1}t_{\rm enc2}$.
Obviously, the times of piercing may lie anywhere during the respective encounters, hence the product 
of the two durations. The factor $\cal N$ is of combinatorial nature. For instance, in $aas$ pairs the 
two antiparallel encounters are indistinguishable such that ${\cal N}_{aas}=2$; 
likewise, ${\cal N}_{api}=2$ since the two stretches of, say, the parallel encounter are indistinguishable; finally, in $ppi$ pairs all 
four orbit stretches (symbolically, $E_1,E_2,E_1,E_2$) are indistinguishable, hence ${\cal N}_{ppi}=4$.

For $\cal T$ invariance, (\r{2}) gives $K^{(3)}_{\rm orth,ind}\!=\!\tau\int_{-c}^c\!du_1\ldots ds_2\\
\f{1}{{\cal N}t_{\rm enc1}t_{\rm enc2}}P(u_1\ldots s_2|T)2\cos{\!\f{\Delta S}{\hbar}}$ where we have
already allowed the HOdA sum rule to yield a factor $\tau$ as above and used the
overcounting factor $\cal N$ as an indicator for the cases $aas,api,ppi$. Using the action difference
(\r{5}) for both encounters and the weight (\r{9}) it is easily found that only the part
$\f{1}{6\Omega^2}\times3\times4\times T^2t_{\rm enc1}t_{\rm enc2}$ in $P$ survives the integration, 
and actually yields the square of the twofold integral met with for $\tau^2$; all other terms and all 
$c$ dependence vanish with $\hbar\to0$. We thus have
\be
\l{10}
K_{\rm orth,ind}^{(3)}=\f{8}{\cal N}\tau^3=\cases{4\tau^3\quad{\rm for}\quad aas,api\cr
                                          2\tau^3\quad{\rm for}\quad ppi\,.\cr}
\ee
For dynamics without $\cal T$ invariance, however, only $ppi$ pairs exist and yield
$K_{{\rm uni,ind}}^{(3)}=\tau^3$, one half the $ppi$ term in (\r{10}) since a $ppi$
orbit now has no time reverse.

{\it $\tau^3$-contributions from triple encounters:} Both for $ac$ and $pc$ pairs, by reasoning as before we have the
density
$$
\nonumber
P^{\rm cl}(u_b,s_b,u_c,s_c|T)\!=\!\f{T\big(T\!-\!3t^{\rm cl}_{\rm enc}\!+\!\dots\big)^2}{2\Omega^2}\!=\!
-\!\f{3T^2t^{\rm cl}_{\rm enc}}{\Omega^2}\ldots 
$$
where the dots point to terms killed by the integration to come, as $\hbar\to0$. The crucial term 
$\propto t_{\rm enc}^{\rm cl}$ is due to minimal loop lengths. The orbit must leave an encounter before reentering in the antiparallel or parallel sense; otherwise, an SR pair already 
accounted for in $\tau^2$ would arise in the antiparallel case, whereas the parallel case would lead 
to a new family with stretches involved in encounters resembling multiple repetitions of shorter 
periodic orbits; such families turn out irrelevant for $\hbar\to0$.  

To count each cloverleaf only once we divide out the familiar ${\cal N}t^{\rm cl}_{\rm enc}$. For 
$pc$ encounters we have ${\cal N}_{pc}=3$ since the three parallel stretches $\triplearrow$ are 
indistingushable while $ac$ encounters $\triplearrow[1]$ entail ${\cal N}_{ac}=1$. Given $\cal T$ 
invariance, the form factor picks up $K^{(3)}_{\rm orth,cl}\!=
\!\tau\!\int_{-c}^cdu_b\ldots ds_c\f{1}{{\cal N}t^{cl}_{\rm enc}}\\P^{\rm cl}(u_b\ldots s_c|T)2
\cos{\!\f{\Delta S}{\hbar}}$. The limit $\hbar\to0$ yields
$$
K^{(3)}_{\rm orth,cl}=-\f{6}{\cal N}\tau^3=\cases{-6\tau^3\quad{\rm for}\quad ac\cr
                                          -2\tau^3\quad{\rm for}\quad pc\,.\cr}
$$
Without $\cal T$ invariance, no $ac$ pairs exist and $pc$ pairs, not\\
accompanied by time inverses, give but $K^{(3)}_{\rm uni,cl}=-\tau^3$.

{\it Conclusions and outlook:} Adding contributions from independent and cloverleaf encounters we get 
$K^{(3)}_{\rm orth}=(4+4+2-6-2)\tau^3=2\tau^3$ and $K^{(3)}_{\rm uni}=0$, as in (\r{1}). To third order in $\tau$ at least, then, semiclassical treatment of {\it individual} hyperbolic dynamics gives the universal form factor characteristic of {\it ensembles of} random matrices.

The five families of orbit pairs met here resemble diagrams known
from field theoretic treatments  of disordered
systems \c{Disorder} and from quantum graphs \c{Graphen}. The analogy between classical orbits and diagrams in field theory should persist in higher orders. Orbit pairs (alias diagrams) with $n\!-\!1$ separate simple encounters contribute to $\tau^n$; upon shrinking intervening loops we expect to find all other relevant orbit pairs. 
The  weight of each family includes a correction, due to the ban of encounter overlap and small as a power of $\f{T_E}{T_H}$, exclusively affecting the form factor. 

Financial support of the Sonderforschungsbereich SFB/TR12 of the Deutsche Forschungsgemeinschaft is 
gratefully acknowledged. We have enjoyed fruitful discussions with A. Altland, G. Berkolaiko, 
K. Richter, M. Sieber, D. Spehner, M. Turek, and M. Zirnbauer.

\end{multicols}

\begin{thebibliography}{9}

\bibitem{BGS} O. Bohigas, M. J. Giannoni, C. Schmit, Phys. Rev. Lett. {\bf 52}, 1 (1984); G. Casati, 
F. Valz-Gris, I. Guarneri, Lett. Nuovo Cim. {\bf 28}, 279 (1980)

\bibitem{Bibel}F. Haake, {\it Quantum Signatures of Chaos}, (Springer, Berlin, 2000)

\bibitem{Gutzi} M. Gutzwiller, {\it Chaos in Classical and Quantum Mechanics}, (Springer, New York 1990)

\bibitem{Smiley} D. Cohen, H. Primack, U. Smilansky, Ann. Phys. {\bf 264}, 108 (1998) and references 
therein

\bibitem{Berry} M. V. Berry, Proc. R. Soc. Lond. {\bf A400}, 229 (1985)

\bibitem{HOdA} J. H.  Hannay, A. M. Ozorio de Almeida, J. Phys. {\bf A17}, 3429 (1984)

\bibitem{Sieber} M. Sieber, K. Richter, Physica Scripta {\bf T90}, 128 (2001); M. Sieber,
J. Phys. {\bf A35}, L613 (2002)

\bibitem{Mueller} S. M\"uller, diploma thesis, Essen (2001); Eur. Phys. J. {\bf B34}, 305 (2003)
\bibitem{Spehner} D. Spehner, J. Phys. {\bf A36}, 7269 (2003)

\bibitem{Regensburg}  M. Turek and K. Richter, J. Phys. {\bf A36}, L455 (2003)

\bibitem{RegEss} S. M\"uller, K. Richter, D. Spehner, M. Turek, to be published 
\bibitem{BHMH} P. Braun, S. Heusler, S. M\"uller, F. Haake, Eur. Phys. J. {\bf B30}, 189 (2002)

\bibitem{BHH} P. Braun, F. Haake, S. Heusler, J. Phys. {\bf A35}, 1381 (2002)

\bibitem{foot} The appearance of $P(u,s|T)/2t_{\rm enc}$ may also be seen as due to  the identity $\int_{-c}^c duds\delta(us-\Delta S)\f{1}{2t_{\rm enc}}P(u,s|T)=
\lambda P\big|_{us=\Delta S}$; note that $u,s$ enter $P(u,s|T)$ and $t_{\rm enc}$  
only through the product $us=\Delta S$. We may interprete $\lambda P\big|_{us=\Delta S}\,d\Delta S$ 
as the number of encounters (thus the number of SR partners) with action difference in 
$[\Delta S,\Delta S+d\Delta S]$, for period-$T$ orbits. Analogous identities hold for the densities met with below for $\tau^3$.

\bibitem{Disorder} R. A. Smith, I. V. Lerner, B. L. Altshuler, Phys. Rev. {\bf B58}, 10343 (1998)

\bibitem{Graphen} G. Berkolaiko, H. Schanz, R. S. Whitney, J. Phys. {\bf A36}, 8373 (2003)
 
\end{thebibliography}
\end{document}